\documentclass[epj]{svjour}

\usepackage{graphics}
\RequirePackage{graphicx}
\RequirePackage{mathptmx}
\RequirePackage[numbers,sort&compress]{natbib}
\RequirePackage{amssymb}
\RequirePackage{bm}
\RequirePackage{slashed}
\usepackage{amssymb}

\newcommand{\be}{\begin{equation}}
\newcommand{\ee}{\end{equation}}
\newcommand{\bea}{\begin{eqnarray}}
\newcommand{\eea}{\end{eqnarray}}
\newcommand{\ba}{\begin{array}}
\newcommand{\ea}{\end{array}}
\newcommand{\la}{\begin{langle}}
\newcommand{\ra}{\begin{rangle}}

\begin{document}

\title{Magnetic moments of the  low-lying ${\frac{1}{2}}^-$ octet baryon resonances}

%\subtitle{Do you have a subtitle?}

\author{Neetika Sharma\inst{1}
        \and
        A. Mart\'inez Torres \inst{2}
        \and
        K. P. Khemchandani \inst{2}
       \and
        Harleen Dahiya \inst{1}
        }

%\thankstext[$\star$]{t1}{Thanks to the title}
%\thankstext{e1}{e-mail: dahiyah@nitj.ac.in}

\institute{Department of Physics, Dr. B.R. Ambedkar National
Institute of Technology, Jalandhar-144011, India.\label{addr1}
          \and
Instituto de F\'isica, Universidade de S\~ao Paulo, C.P 66318, 05314-970 S\~ao Paulo, SP, Brazil.\
           \label{addr2}}

\mail{dahiyah@nitj.ac.in}
\date{Received: date / Accepted: date}

%\maketitle

\abstract{
The magnetic moments of the negative parity octet resonances  with spin ${\frac{1}{2}}$: $N^*$(1535), $N^*$(1650), $\Sigma^*$(1620), and $\Xi^*$(1690) have been calculated
within the framework of the chiral constituent quark model.
In this approach, the presence of the polarized $q\bar{q}$ pairs (or the meson cloud, in other words) is considered by
using the Lagrangian for Goldstone boson emission from the constituent quarks. Further, the explicit contributions coming from the spin and orbital angular momentum,  including the effects of the configurations mixing
between the states with different spins,
are obtained. The motivation for  these calculations comes from the recent interest in experimental measurement of the magnetic moment
of the ${S_{11}(1535)}$ resonance and of similar calculations being done within lattice quantum chromodynamics approaches.
Our results can be compared with  those expected to come from these sources.}

\maketitle

\section{Introduction}

Scrutinizing the structure and the properties of light hadrons is a key to understanding the mechanism of the strong interactions at low energies. The very feature of the confinement of the theory of the strong interactions makes it difficult to access the underlying physics at the low and intermediate energies. However, continuous efforts are being made to investigate this energy region and a lot of important information is being made available through theoretical and experimental investigations.
Studies of  hadrons within constituent quark models have played a very important role in some crucial discoveries like those of the color quantum number and  the charm flavor of the quarks. This model was later used to determine the masses of the excited states in the baryon spectrum by replacing each quark by  a harmonic oscillator \cite{ik771,ik772,ik773,ik78}.
The n\"aive quark model has been further extended by implementing the chiral symmetry and its spontaneous breaking to it  \cite{chiral}, giving rise to the chiral constituent quark model ($\chi$CQM), which as a result includes the important phenomenon  of quark-antiquark excitations, in other words the presence of the meson cloud at low energies. This perspective is in common with the modern effective field theory approaches used to study the baryon resonances \cite{jido20021,jido20022,hyodo,detar,jido,Geng11,Geng12,Geng13,three}.

%In the frame of constituent quark model, a baryon resonance is described as an
%excited state of a nucleon where three constituent quarks interact by exchanging
%gluons or Goldstone Bosons in a confining potential. In fact, there are two kinds of
%exchanges. Some models, such as QCD inspired Isgur-Karl model \cite{ik77,ik78} and
%flux-tube model \cite{paton}, assume that all spin dependencies arise from one gluon
%exchange. Other models, such as Goldstone Boson exchange model (GBE) \cite{gbe}, believe
%that the short-distance force between quarks is mediated by the exchange of nearly massless
%Goldstone Bosons generated by the spontaneous breaking of chiral symmetry. The GBE (Goldstone-
%boson-exchange model) are based upon the Hamiltonian approach with harmonic-oscillator wave
%functions whereas the chiral constituent quark model, assume the short-distance force between
%quarks is mediated by the exchange of nearly massless Goldstone bosons GBs generated by the
%spontaneous breaking of chiral symmetry Lagrangian.
%

The $\chi$CQM, which facilitates the calculations of hadron properties from a theory based on the symmetries of the quantum chromodynamics (QCD), successfully explains the ``Proton spin crisis'' and other related properties \cite{cheng1,cheng2,johan1,johan2,johan3,johan4,hd1,hd2,hd3,hd4,hd5,nssigma1,nssigma2,nssigma3,nssigma4}. Later, the octet and decuplet baryon magnetic moments were calculated by incorporating the sea quark polarizations and their orbital angular momentum contribution through  the generalized Cheng-Li mechanism \cite{chengmag1,chengmag2,chengmag3}. Recently, the model has been used to study the charm sector too \cite{nscharm}.
Apart from studying the static properties of the ground state baryons, the $\chi$CQM has also been found useful to obtain the $D$/$F$ ratio for the axial vector coupling constants, radiative decays, etc..

Considering the success of the $\chi$CQM in studying ground state baryons, it is natural to extend this model to investigate the properties of the excited baryons. In fact the quark model and $\chi$CQM  have already been used to study the magnetic moments \cite{chiang,jliu}, radiative decays \cite{raddecay} and dynamical properties \cite{seven} of some baryon resonances, and a reasonable agreement with the experimental data, where it exists, has been found. In this work, we use this model to investigate the magnetic moments of the low-lying, spin-parity $J^P={\frac{1}{2}}^-$ octet resonances.

The present work is  partly motivated by the  interest in the experimental studies of the magnetic moment of the $N^*$ planned at the
Mainz Microtron (MAMI) facility \cite{kotulla1,kotulla2,susan}, by the preliminary calculations done with lattice QCD \cite{lattice} and partly by the other calculations of the same done within very different theoretical models \cite{jido20021,jido20022,hyodo}.  Our present work assumes that the basic properties of the baryon resonances can be explained in terms of  three valence quarks surrounded by the meson cloud. However, baryon resonances could possess a different structure, for instance, some of them have been claimed to be multiquark states \cite{eight}, others as dynamically generated ones \cite{jido20021,jido20022,hyodo,six1,six2,five1,five2,nine1,nine2,ninepr}, yet others as hybrids \cite{hybrid1,hybrid2,hybrid3}, etc.  Eventually, the claimed nature of these states can be verified by comparing the model calculations of observables with the experimental data. One of the observables useful for such purposes  is the magnetic moment, which is what we are concerned with here.

The question now comes if the magnetic moments of all the baryon resonances  studied here can be measured.
The number of experiments planned to obtain these observables are rather scarce \cite{kotulla1,kotulla2,susan} as compared to the number of known baryon resonances \cite{pdg}.
The main obstacle in carrying out such experiments  is the considerable widths of excited baryons, which is
of the order of 100 MeV.  Due to this the magnetic moments have to be extracted from the polarization observables
of the decay products of the resonances excited in photo- or pion-induced reactions, and the procedure finally depends on model calculations to estimate
the background. Nevertheless, recent efforts put in measuring
the magnetic moment of $\Delta^+$(1232), through the  $\gamma p \to  \Delta^+ \to p \pi^0 \gamma^\prime$ reaction,  and
the  plan to obtain this information for  $N^*$(1535),  by studying  the $\gamma p \to  p \eta \gamma^\prime$ process, at the MAMI facility, the measurements
of form factors of resonances at the Thomas Jefferson National Accelerator Facility (JLAB) \cite{jlab11,jlab12,jlab13,jlab14}, show that
the interest in this field is growing.

To do the present calculation, we take the benefit from the earlier works  \cite{ik771,ik772,ik773,ik78} where the spectrum for the negative parity
partners of the octet baryons with spin ${\frac{1}{2}}$ was obtained  in a quark model, by solving a bound state equation treating the three quarks as harmonic oscillators,  considering
excitation of one unit of the orbital angular momentum, and the resulting states were found to be in good agreement with the known ones. The starting point of
our model is the spectrum found in  Refs.~\cite{ik771,ik772,ik773,ik78}, which provides  the wave functions of the ${\frac{1}{2}}^-$ baryon resonances needed to calculate the magnetic moments.
From the quark model studies, it is known that the magnetic moment of  the low-lying spin ${\frac{1}{2}}^+$ baryons receives contributions not only from the valence quarks, but also from many other effects, such as the
quark sea, the orbital angular momentum,  relativistic effects, the meson
cloud effect, etc. \cite{val1,val2,val3,val4,val5,val6,val7,chengmag1,chengmag2,chengmag3}. In the present work, we will calculate all these contributions for  their low-lying spin ${\frac{1}{2}}^-$ counterparts.

The paper is organized as follows: we first formulate in detail the
magnetic moments of the negative parity $N^*$
resonances ($S_{11}^+(1535)$, $S_{11}^0(1535)$,
$S_{11}^+(1650)$, and $S_{11}^0(1650)$). In Sec. \ref{modelmdm}, the explicit contribution of the spin and orbital angular momentum of the magnetic moment for the above mentioned resonances  are calculated  in the nonrelativistic quark model.
Further, we calculate explicitly the
contribution coming from the valence quarks,  and quark sea
polarization, including the Cheng-Li mechanism, for the spin part and the valence and the sea quarks contribution for the orbital angular momentum part in the $\chi$CQM. The details of these calculations have been presented in Sec. \ref{cqm}.  In
Sec. \ref{result} we discuss our numerical results  for all the ${\frac{1}{2}}^-$ octet  baryon resonances and Sec. \ref{summary} summarizes our results.

\section{Magnetic moment of baryons in the quark model} \label{modelmdm}

In the nonrelativistic SU(6) constituent quark model (NCQM), the magnetic moment of the baryon resonances have contribution coming from both quark spin and orbital
angular momentum \be  \bm{\mu}_{B} = \bm{\mu^S}_{B} + \bm{\mu^L}_{B} \,, \label{spinorbit} \ee with
\bea \label{eq:}
\bm{\mu^S}_{B}
&=& \sum_i \bm{\mu^s}_i
 =  \sum_i \frac{Q_i}{ m_i}\, \bm{s}_i \,, \label{spin1}\\
\bm{\mu^L}_{B}
&=& \sum_i \bm{\mu^l}_i
=  \sum_i \frac{Q_i}{2m_i}\, \bm{l}_i \,,
\label{orbit1}
\eea
where ${\bf s_i}$ and ${\bf l_i}$ are the spin and orbital angular momentum of
the $i$th quark and the index $i$ is summed over three quarks. The orbital angular momentum part vanishes for ground state
baryons on account of the absence of orbital excitation. If $|B \rangle$ is the given baryon's SU(6) spin-flavor wavefunction, then
we have \bea \langle B|\mu^S|B\rangle  &=& \Delta u \mu_u +\Delta
d \mu_d +\Delta s \mu_s\,, \label{spin2} \\ \langle B|\mu^L|B\rangle  &=& \Delta
u^{(1)} \mu_u +\Delta d^{(1)} \mu_d +\Delta s^{(1)} \mu_s \label{orbit2} \,.\eea
Here, $\mu_q = \frac{e_q}{2 M_q}$
($q=u,d,s$) is the quark magnetic moment, $e_q$ and $M_q$ are the
electric charge and mass of the $q$ quark, respectively.
The polarizations corresponding to the spin and orbital angular momentum, $ \Delta q = q^{\uparrow} - q^{\downarrow}$
and $ \Delta q^{(1)} = q^{(1)} - q^{(-1)}$, can be calculated as
\be\widehat B \equiv \langle B|{\cal
N} |B \rangle = \langle B|{\cal
N}^{\uparrow\downarrow} + {\cal N}^{(\pm 1)} |B \rangle.\, \label{bnb1} \ee
In Eq.~(\ref{bnb1}),  ${\cal N^{\uparrow\downarrow}}$ and ${{\cal N}}^{(\pm 1)}$ are the number
operators given by
\bea {\cal N^{\uparrow\downarrow}} &=&
n_{u^{\uparrow}}u^{\uparrow} + n_{u^{\downarrow}}u^{\downarrow} +
n_{d^{\uparrow}}d^{\uparrow} + n_{d^{\downarrow}}d^{\downarrow} +
n_{s^{\uparrow}}s^{\uparrow} + n_{s^{\downarrow}}s^{\downarrow} \,,
%\label{number1}
\\  {{\cal N}}^{(\pm 1)} &=& n_{u^{(1)}}u^{(1)}+
n_{u^{(-1)}}u^{(-1)} + n_{d^{(1)}}d^{(1)} + n_{d^{(-1)}}d^{(-1)} \nonumber \\&& +
n_{s^{(1)}}s^{(1)} + n_{s^{(-1)}}s^{(-1)} \,, \label{number2}\eea
with $n_{q^{\uparrow}}$ ($n_{q^{\downarrow}}$)
being the number of quarks with spin up (down) and $n_{q^{(1)}}$ ($n_{q^{(-1)}}$) is the number of quarks
with the orbital angular momentum projection $m_L=1$ ($m_L=-1$).

In this section, we calculate the magnetic moment of
the $S_{11}(1535)$ and $S_{11}(1650)$ nucleon resonances in the framework of the NCQM
and further extend this to the rest of low-lying negative parity octet baryons. The lowest lying
negative parity nucleon resonances are $|N^2P_{1/2} \rangle$ and
$|N^4P_{1/2} \rangle$, where the usual spectroscopic notation $^{2S +1 } L_J$ is
used to indicate their total
quark spin $S={\frac{1}{2}},\,{\frac{3}{2}}$ ($2S+1=2,\,4$), orbital angular momentum
$L=1$ ($P$-wave), and total angular momentum $J={\frac{1}{2}}$ \cite{ksw95}.
The spin angular momentum $S={\frac{1}{2}}$ couples with the orbital angular
momentum $L=1$ to give the total angular momentum $J={\frac{1}{2}}$ and
$J={\frac{3}{2}}$.
%In this work, we calculate the magnetic moment of the
%states with $J=1/2$ only.
The wavefunctions of the $|N^2P_{1/2} \rangle$
and $|N^4P_{1/2} \rangle$ states are explicitly given as \bea
\label{eq:N24} |{N^2P_{1/2}} \rangle &= &  \frac{1}{\sqrt{2}}
\sum_{m_l m_s} {\langle{ 1~\frac{1}{2}~\ m_l~m_s|
\frac{1}{2}~\frac{1}{2}}\rangle } \biggl\{ \psi^\rho_{1m_l}
\Bigl[ \frac{1}{\sqrt{2}} \bigl( \chi^\lambda_{m_s}\,\phi^\rho \nonumber \\  && +
\chi^\rho_{m_s}\,\phi^\lambda \bigr) \Bigr]  +
\psi^\lambda_{1m_l} \Bigl[ \frac{1}{\sqrt{2}} \bigl(
\chi^\rho_{m_s}\,\phi^\rho - \chi^\lambda_{m_s}\,\phi^\lambda \bigr) \Bigr] \biggr\} \,, \\
| N^4P_{1/2} \rangle & =& \frac{1}{\sqrt{2}} \sum_{m_l m_s}
\langle\,1\,\frac{3}{2}\,m_l\ m_s \,|\, \frac{1}{2}\, \frac{1}{2}
\,\rangle \Bigl[\,\psi^\rho_{1m_l}\,\chi^s_{m_s}\,\phi^\rho + \nonumber \\  &&
\,\psi^\lambda_{1m_l}\,\chi^s_{m_s}\,\phi^\lambda\,\Bigr] \,, \eea
where $\psi$, $\chi$, and $\phi$ denote the spatial, spin, and
flavor wavefunctions. The superscripts $s$ or $\rho$ ($\lambda$)
indicate that they are totally symmetric among three quarks, or odd
(even) under the exchange of the first two quarks \cite{ik771,ik772,ik773}.

The physical eigenstates for the $L=1$
negative parity resonances, are linear combinations of these two and are expressed as \bea |S_{11}(1535) \rangle = \cos \theta|N^2P_{1/2}
\rangle -\sin \theta|N^4P_{1/2}\rangle
 \,, \\
|S_{11}(1650) \rangle= \sin\theta|N^2P_{1/2}\rangle + \cos\theta|N^4P_{1/2}\rangle\,.
\eea
Now, the magnetic moments of the $S_{11}(1535)$ and $S_{11}(1650)$
resonances can be expressed in terms of the magnetic moments of the
$|N^2P_{1/2}\rangle$ and $|N^4P_{1/2}\rangle$ states and their cross terms as
\bea \label{eq:mus10a}
\mu_{S_{11}(1535)}
&=& \mu_{N^2P_{1/2}} \cos^2\theta + \mu_{N^4P_{1/2}} \sin^2\theta -
\nonumber \\  && 2\,\langle\, N^2P_{1/2} \,|\mu_z|\, N^4P_{1/2} \,\rangle
 \cos\theta \sin\theta \,,
\\
\label{eq:mus10b}
\mu_{S_{11}(1650)}
  &=& \mu_{N^2P_{1/2}} \sin^2\theta + \mu_{N^4P_{1/2}} \cos^2\theta +
 \nonumber \\  &&  2\,\langle\, N^2P_{1/2} \,|\mu_z|\, N^4P_{1/2}\,\rangle
  \cos\theta \sin\theta \,.
\eea
%%The value of the mixing angle $\theta$ depends on the quark interaction.
%%change
%The value of mixing angles of the wave functions of the
%negative-parity low-lying nucleon resonances are sensitive to the different models with different
%hyperfine interactions \cite{gloz,gbe}.
The value of the mixing angle $\theta$ depends on the quark interaction and predictions for it are available from different models. For example, in Ref.~\cite{ik771,ik772,ik773} it was considered that the splitting among the negative-parity baryons  is dominantly explained in terms of an interaction whose spin dependent part  originates from one gluon exchange between the quarks. Within this formalism, the authors predicted  a mixing angle of $\theta =\tan^{-1} {(\sqrt{5} -1)/2} \simeq -31.7^\circ$ for the $J=1/2$ case. This value is close to the empirical mixing angle $\theta \simeq -32^\circ$ found in Ref. \cite{ajhey}. In other models~\cite{Glozman:1995fu,Wagenbrunn:2000kr,Garcilazo:2001md}, the mixing angle present in the wave function of the resonances is determined considering the spontaneous breaking of chiral symmetry and the Goldstone boson exchange between quarks. In this case, a mixing angle of $-27\pm 12$ is found. Both models are known to well describe the spectrum of the $1/2^-$ low-lying baryon resonances. In this work, we have used the value of the mixing angle obtained in Ref.~\cite{ik771,ik772,ik773}. For a  compilation of the results obtained by using mixing angles given by different models  we refer the reader to, for example, Refs.~\cite{He:2003vi1,He:2003vi2,He:2003vi3}.

Since the magnetic moment has contribution coming from both spin and orbital angular momentum as expressed in Eqs.~(\ref{spin1}) and (\ref{orbit1}), we now present the matrix elements for quark spin and orbital
angular momentum contributions of the magnetic moments of the states $|N^2
P_{1/2}\rangle$ and $|N^4 P_{1/2}\rangle$ and the cross terms obtained from
mixing the $|N^2P_{1/2}\rangle$ and $|N^4P_{1/2}\rangle$ states.

It is easy to obtain the spin structure of the baryons as expressed in Eq.~(\ref{bnb1}) \bea \langle
N^2P_{1/2}^+ |{\cal N}| N^2P_{1/2}^+ \rangle &=&
\frac{8}{9}u^{\uparrow} +\frac{10}{9} u^{\downarrow}
+\frac{4}{9}d^{\uparrow} +\frac{5}{9} d^{\downarrow} \nonumber \\ &&
+\frac{4}{9}u^{(1)} +\frac{2}{9}d^{(1)} \,, \nonumber \\
\langle
N^2P_{1/2}^0 |{\cal N}| N^2P_{1/2}^0 \rangle &=&
\frac{4}{9}u^{\uparrow} +\frac{5}{9} u^{\downarrow}
+\frac{8}{9}d^{\uparrow} +\frac{10}{9} d^{\downarrow} \nonumber \\ &&
+\frac{2}{9}u^{(1)} +\frac{4}{9}d^{(1)} \,, \nonumber \\
\langle
N^4P_{1/2}^+|{\cal N}|N^4P_{1/2}^+ \rangle &=&
\frac{14}{9}u^{\uparrow} +\frac{4}{9}u^{\downarrow}
+\frac{7}{9}d^{\uparrow}+\frac{2}{9}d^{\downarrow}+\nonumber \\ &&  \frac{1}{18}u^{(1)} + \frac{1}{6}u^{(-1)}  +\frac{1}{9}d^{(1)}
+\frac{1}{3}d^{(-1)}\,, \nonumber \\
\langle
N^4P_{1/2}^0|{\cal N}|N^4P_{1/2}^0 \rangle &=&
\frac{7}{9}u^{\uparrow} +\frac{2}{9}u^{\downarrow}
+\frac{14}{9}d^{\uparrow}+\frac{4}{9}d^{\downarrow}+ \nonumber \\ && \frac{1}{9}u^{(1)} + \frac{1}{3}u^{(-1)}  +\frac{1}{18}d^{(1)}
+\frac{1}{6}d^{(-1)}\,, \nonumber \\ \langle N^2P_{1/2}^+|{\cal N}
| N^4P_{1/2}^+ \rangle &=&
\frac{2}{9}u^{\uparrow} -
\frac{2}{9}u^{\downarrow} -\frac{2}{9}d^{\uparrow}
+\frac{2}{9}d^{\downarrow} \,, \nonumber \\
\langle N^2P_{1/2}^0|{\cal N}
| N^4P_{1/2}^0 \rangle &=&
-\frac{2}{9}u^{\uparrow} +
\frac{2}{9}u^{\downarrow} +\frac{2}{9}d^{\uparrow}
-\frac{2}{9}d^{\downarrow} \,.
\eea
Here, the superscripts + and 0 denote the
charge of the resonance state. It is important to mention at this point that there are no cross terms for $\mu^L_z$ because
$|N^2P_{1/2}\rangle$ and $|N^4P_{1/2}\rangle$ have orthogonal quark spin
states which are not affected by $\mu^L_z$.
Using  Eqs.~(\ref{spin2})  and (\ref{orbit2}) the magnetic moment of the
$|N^2P_{1/2}^+ \rangle$, $|N^4P_{1/2}^+ \rangle$, and the cross terms can be expressed as
\bea
 \label{eq:mu1}
&&\mu_{N^2P_{1/2}^+} = -\frac{2}{9} \mu_u - \frac{1}{9}\mu_d
+\frac{4}{9} \mu_u + \frac{2}{9}\mu_d \,,
\\ \label{eq:mu2}  &&\mu_{N^2P_{1/2}^{0}} = -\frac{1}{9}\mu_u-
\frac{2}{9}\mu_d +\frac{2}{9}\mu_u+ \frac{4}{9}\mu_d\,,
\\
\label{eq:mu3}
&&\mu_{N^4P_{1/2}^{+}} = \frac{10}{9}\mu_u+ \frac{5}{9}\mu_d -
\frac{1}{9}\mu_u - \frac{2}{9}\mu_d\,,\\
\label{eq:mu4}
&&\mu_{N^4P_{1/2}^{0}} = \frac{5}{9}\mu_u + \frac{10}{9} \mu_d-
\frac{2}{9}\mu_u - \frac{1}{9}\mu_d \,.\\
&&\langle N^4P_{1/2}^+|\mu^S_z |N^2P_{1/2}^+\rangle
= \frac{4}{9}\mu_u - \frac{4}{9}\mu_d \,, \label{eq:mu5}\\
&&\langle\, N^4P_{1/2}^{0}|\mu^S_z|N^2P_{1/2}^{0}\rangle
= -\frac{4}{9}\mu_u + \frac{4}{9} \mu_d \label{eq:mu6}
\,. \eea

From Eqs.~(\ref{eq:mus10a}) and (\ref{eq:mus10b}), the magnetic moment of the $S_{11}(1535)$ and $S_{11}(1650)$ states
in the NCQM are expressed as
\bea
\label{eq:mus11a}
\mu_{S_{11}^+(1535)} &=& \left(-\frac{2}{9}\mu_u
-\frac{1}{9}\mu_d +\frac{4}{9}\mu_u +\frac{2}{9}\mu_d \right)
\cos^2\theta +\nonumber \\ &&  \left(\frac{10}{9}\mu_u +\frac{5}{9}\mu_d -\frac{1}{9}\mu_u -\frac{2}{9}\mu_d \right)
\sin^2\theta - \nonumber \\ && \left( \frac{8}{9}\mu_u
-\frac{8}{9}\mu_d \right) \cos\theta \sin\theta  \,,\\
\label{eq:mus11b}
\mu_{S_{11}^0(1535)} &=& \left(-\frac{1}{9}\mu_u -\frac{2}{9}\mu_d
+\frac{2}{9}\mu_u +\frac{4}{9}\mu_d \right) \cos^2 \theta  +\nonumber \\ &&
\left(\frac{5}{9}\mu_u +\frac{10}{9}\mu_d -\frac{2}{9}\mu_u -\frac{1}{9}\mu_d \right) \sin^2\theta - \nonumber \\ &&
 \left(-\frac{8}{9}\mu_u + \frac{8}{9}\mu_d
\right) \cos\theta \sin\theta \,,\\
\label{eq:mus11c}
\mu_{S_{11}^+(1650)} &=&
\left(-\frac{2}{9}\mu_u -\frac{1}{9}\mu_d +\frac{4}{9}\mu_u
+\frac{2}{9}\mu_d \right) \sin^2\theta  + \nonumber \\ && \left(\frac{10}{9}\mu_u
+\frac{5}{9}\mu_d -\frac{1}{9}\mu_u
-\frac{2}{9}\mu_d \right) \cos^2\theta + \nonumber \\ &&
\left(\frac{8}{9}\mu_u -\frac{8}{9}\mu_d \right) \cos\theta \sin\theta \,,\\
\label{eq:mus11d}
\mu_{S_{11}^0(1650)} &=&
\left(-\frac{1}{9}\mu_u
-\frac{2}{9}\mu_d +\frac{2}{9}\mu_u +\frac{4}{9}\mu_d \right) \sin^2 \theta  + \nonumber \\ && \left(\frac{5}{9}\mu_u +\frac{10}{9}\mu_d -\frac{2}{9}\mu_u  -\frac{1}{9}\mu_d \right)
\cos^2\theta  + \nonumber \\ &&   \left(-\frac{8}{9}\mu_u
+\frac{8}{9}\mu_d \right) \cos\theta \sin\theta \,.\eea

Considering
$m_u=m_d=\frac{1}{3}m_N$, we have  $\mu_u=Q_u/2m_u=2 \mu_N$ and
$\mu_d=Q_d/2m_d= -\mu_N$. Thus, substituting these values in
Eqs.~(\ref{eq:mus11a})-(\ref{eq:mus11d}), we obtain
\bea
\label{eq:mus2a}
\mu_{S_{11}^+(1535)} &=& \frac{1}{3} \cos^2 \theta +\frac{5}{3} \sin^2 \theta - \frac{8}{3} \cos \theta \sin \theta \,,\\
\label{eq:mus2b}
\mu_{S_{11}^0(1535)} &=& -\frac{1}{3} \sin^2 \theta + \frac{8}{3} \cos \theta \sin \theta\, , \\
\label{eq:mus2c}
\mu_{S_{11}^+(1651)} &=& \frac{5}{3} \cos^2 \theta +\frac{1}{3} \sin^2 \theta + \frac{8}{3} \cos \theta \sin \theta\, , \\
\label{eq:mus2d}
\mu_{S_{11}^0(1651)} &=&  -\frac{1}{3} \cos^2 \theta  - \frac{8}{3} \cos \theta \sin \theta \,. \eea
Using the value of mixing angle $\theta = -31.7^\circ$
in
Eqs.~(\ref{eq:mus2a})-(\ref{eq:mus2d}), one can obtain the magnetic moments of the $N^*$ resonances $S_{11}^+(1535)$, $S_{11}^0(1535)$, $S_{11}^+(1650)$, and $S_{11}^0(1650)$. The magnetic moment of the other negative parity low-lying baryon resonances with spin ${\frac{1}{2}}$ can similarly be calculated using their respective wave functions. The results of these calculations are presented and discussed in Section~\ref{result}.

\section{Magnetic moment of baryons in $\chi$CQM}
\label{cqm}
The key to understanding the magnetic moment of the baryons
in the $\chi$CQM formalism \cite{cheng1,cheng2} is the fluctuation process \be
q^{\uparrow\downarrow} \rightarrow {\rm GB} + q^{' \downarrow\uparrow}
\rightarrow (q \bar q^{'})
+q^{'\downarrow\uparrow}\,, \label{basic}\ee where GB represents the
emitted Goldstone boson and $q \bar
q^{'} +q^{'}$ constitute the ``quark sea'' \cite{hd1,hd2,hd3,hd4,hd5,johan1,johan2,johan3,johan4}.
The model assumes that the quark flavor, spin and orbital contents of the baryons are determined by its valence quark structure and all possible chiral fluctuations, and probabilities of these fluctuations depend on the interaction Lagrangian,
\begin{equation}
 {\cal L}= g_8 {\bf \bar
q}\left(\Phi+\zeta\frac{\eta'}{\sqrt 3}I \right) {\bf q}=g_8 {\bf
\bar q}\left(\Phi' \right) {\bf q}\,,\label{lag}
\end{equation}
where the coupling between the quark and Goldstone bosons is weak enough to treat the fluctuation (\ref{basic}) as a small perturbation and the contributions from the higher order fluctuations are assumed to be negligible.  Clearly, the GB emission from the quark gives rise to the probability of  a spin-flip of the quark. It is assumed that the spin-flip process gives the dominant contribution to (\ref{basic}), which is similar to the chiral instanton model \cite{instanton1,instanton2}.
In Eq.~(\ref{lag}),
$\zeta=g_1/g_8$, $g_1$ and $g_8$ denote the coupling constants for the
singlet and octet GBs, respectively, and $I$ is the $3\times 3$ identity
matrix. The matrix of the GBs can
be expressed as
\bea \Phi' =\left(
\ba{ccc} \frac{\pi^0}{\sqrt 2} + \beta\frac{\eta}{\sqrt 6} +
\zeta\frac{\eta^{'}}{\sqrt 3} & \pi^+& \alpha K^+ \\ \pi^- &
-\frac{\pi^0}{\sqrt 2} +\beta \frac{\eta}{\sqrt 6}
+\zeta\frac{\eta^{'}}{\sqrt 3} & \alpha K^0
\\\alpha K^-  & \alpha \bar{K}^0 & -\beta \frac{2\eta}{\sqrt 6}
+\zeta\frac{\eta^{'}}{\sqrt 3} \ea \right) \nonumber \eea
\be ~~~{\rm and} ~~~~ q
= \left( \ba{c} u \\ d \\ s \ea \right)\, \ee
The parameter
$a(=|g_8|^2$) denotes the probability of chiral fluctuation  $u(d)
\rightarrow d(u) + \pi^{+(-)}$, whereas $\alpha^2 a$, $\beta^2 a$
and $\zeta^2 a$ respectively denote the probabilities of
fluctuations $u(d) \rightarrow s + K^{-(0)}$, $u(d,s) \rightarrow
u(d,s) + \eta$, and $u(d,s) \rightarrow u(d,s) + \eta^{'}$.

The spin part $\mu^S$ of the magnetic moment of a given baryon receives
contributions from
the valence quarks, sea quarks, and orbital angular momentum of the
``quark sea'' \cite{chengmag1,chengmag2,chengmag3} and is expressed as \bea
\mu^S_{B} &=& \mu^S_{{\rm val}} + \mu^S_{{\rm sea}} + \mu^S_{{\rm
orbit}} \,,\label{totmag1} \eea
where $\mu^S_{{\rm val}}$ and $\mu^S_{{\rm sea}}$  represent the
contributions of the valence
quarks and the sea quarks to the magnetic moments due to the spin
polarizations. In addition,
there is a significant contribution coming from the
orbital angular momentum of the ``quark sea'',
$\mu^{S}_{\textrm{orbit}}$, since the GB
emitted
due to the chiral fluctuation is in the P wave state, $\langle l_z
\rangle=1$.  The details
of the valence quark calculations have already been presented in the
previous section. We now present the calculations of the sea and
orbital angular momentum contributions.

The sea quark spin polarizations corresponding to each baryon can be
obtained by substituting for each valence quark
\be
q^{\uparrow(\downarrow)} \rightarrow   -P_{[q, ~GB]}
q^{\uparrow(\downarrow)}+ |\psi(q^{\uparrow(\downarrow)})|^2\,, \label{qp1}
\ee
when calculating the spin
contribution to the magnetic moment. In Eq.~(\ref{qp1}),
$P_{[q, ~GB]}$ is the probability of emission of GBs
from a quark $q^{\uparrow(\downarrow)}$ and
$|\psi(q^{\uparrow(\downarrow)})|^2$ is the probability
of transforming a $q^{\uparrow(\downarrow)}$ quark \cite{hd1,hd2,hd3,hd4,hd5} given by \bea
|\psi(u^{\uparrow(\downarrow)})|^2 &=&  \frac{a}{6}\left(3 +
\beta^2 + 2 \zeta^2 \right)u^{\downarrow(\uparrow)}+ a d^{\downarrow(\uparrow)} + a \alpha^2
s^{\downarrow(\uparrow)}\,, \nonumber \\
|\psi(d^{\uparrow(\downarrow)})|^2 &=&  a u^{\downarrow(\uparrow)}+ \frac{a}{6}
\left(3+\beta^2+2 \zeta^2 \right)d^{\downarrow(\uparrow)}+
a \alpha^2 s^{\downarrow(\uparrow)} \,, \nonumber\\ |\psi(s^{\uparrow(\downarrow)})|^2 &=& a
\alpha^2 u^{\downarrow(\uparrow)} + a\alpha^2 d^{\downarrow(\uparrow)} +
\frac{a}{3} \left(2 \beta^2 + \zeta^2 \right)s^{\downarrow(\uparrow)} \,,
\label{psiup} \eea
for the spin up quarks.

The orbital angular momentum contribution of each chiral
fluctuation is given as \cite{chengmag1,chengmag2,chengmag3,cheng1,cheng2} \be \mu(q^{\uparrow}
\rightarrow {q}^{'\downarrow}) = \frac{e_{q^{'}}}{2M_q} \langle l_q
\rangle + \frac{{e}_{q} - {e}_{q^{'}}}{2 {M}_{{\rm GB}}}\langle
{l}_{{\rm GB}} \rangle \label{qq} \,, \ee where $\langle l_q \rangle
= \frac{{M}_{{\rm GB}}}{M_q + {M}_{{\rm GB}}}$ and $\langle l_{{\rm
GB}} \rangle = \frac{M_q}{M_q + {M}_{{\rm GB}}} \label{lq}$. The
quantities ($l_q$, $l_{{\rm GB}}$) and ($M_q$, ${M}_{{\rm GB}}$) are
the orbital angular momenta and masses of the quarks and GBs,
respectively. The orbital moment of each process in Eq.~(\ref{qq}) is
then multiplied by the probability for such a process to take place
to yield the magnetic moment due to all the transitions starting
with a given valence quark. For example, \bea [\mu(u_{\uparrow}
\rightarrow )] &=& a \left [ \left(\frac{1}{2} +\frac{\beta^2}{6}+
\frac{\zeta^2}{3} \right) \mu (u_{\uparrow} \rightarrow u_\downarrow) + \mu
(u_{\uparrow}\rightarrow d_\downarrow) \right. \nonumber \\ && \left.  + \alpha^2 \mu (u_\uparrow
\rightarrow s_\downarrow)\right ]
\,, \label{muu} \eea
\bea [\mu (d_{\uparrow} \rightarrow )] &=& a \left [
\mu (d_{\uparrow} \rightarrow u_\downarrow)+ \left (\frac{1}{2}+
\frac{\beta^2}{6} +
\frac{\zeta^2}{3} \right) \mu (d_{\uparrow} \rightarrow
d_\downarrow) \right. \nonumber \\ && \left. + \alpha^2 \mu
(d_\uparrow \rightarrow s_\downarrow) \right ] \,, \label{mud} \eea \bea
[\mu (s_{\uparrow}
\rightarrow )] &=& a \left [\alpha^2 \mu (s_{\uparrow} \rightarrow u_\downarrow) +
\alpha^2 \mu (s_\uparrow \rightarrow d_\downarrow) \right. \nonumber \\ && \left. + \left (\frac{2}{3} \beta^2+
\frac{\zeta^2}{3} \right) \mu (s_{\uparrow} \rightarrow s_\downarrow )
\right ] \,.
\label{mus} \eea  The orbital
moments of the $u$, $d$, and $s$ quarks in terms of
the $\chi$CQM
parameters ($a, \alpha, \beta, \zeta$), quark masses ($M_u,M_d,M_s$)
and GB masses ($M_{\pi},M_{k},M_{\eta},M_{\eta'}$), are respectively
given as \bea [\mu(u_\uparrow \rightarrow)] &=& a \left [\frac{3 M^2_{u}}{2
{M}_{\pi}(M_u+ {M}_{\pi})}- \frac{\alpha^2(M^2_{K}- 3 M^2_{u})}{2
{M}_{K}(M_u+ {M}_{K})} \right. \nonumber \\ && \left. + \frac{\beta^2 M_{\eta}}{6(M_u+
{M}_{\eta})}+ \frac{\zeta^2 M_{\eta'}}{3(M_u+ {M}_{\eta'})}  \right]
{\mu}_u \,, \label{orbitu} \eea
\bea [\mu(d_\uparrow \rightarrow)] &=& -2 a
\left [\frac{3( M^2_{\pi}-2 M^2_{d})}{4 {M}_{\pi}(M_d+ {M}_{\pi})}-
\frac{\alpha^2 M_{K}}{2(M_d+ {M}_{K})}\right. \nonumber \\ && \left.  - \frac{\beta^2
M_{\eta}}{12(M_d+ {M}_{\eta})}- \frac{\zeta^2 M_{\eta'}}{6(M_d+
{M}_{\eta'})} \right ] {\mu}_d \,, \label{orbitd} \eea \bea [\mu(s_\uparrow
\rightarrow)] &=& -2 a \left[ \frac{\alpha^2 (M^2_{K}-3
M^2_s)}{2{M}_{K}(M_s+ {M}_{K})}  - \frac{\beta^2 M_{\eta}}{3(M_s+
{M}_{\eta})} \right. \nonumber \\ && \left. - \frac{\zeta^2 M_{\eta'}}{6(M_s+ {M}_{\eta'})} \right
]{\mu}_s\,. \label{orbits} \eea

Using this formalism, we can calculate explicitly the valence,
sea, and orbital contributions to the spin angular momentum of
the magnetic moments of the baryons. As an example,
the $\mu^S$ contribution to the magnetic moment of the
$S_{11}^+(1535)$ is given as
\bea \mu^S_{{\rm val}}(S_{11}^+(1535)) &=& -\left(\frac{2}{9}\mu_u
+\frac{1}{9}\mu_d \right)\cos^2 \theta  + \left(\frac{10}{9}\mu_u
+ \right. \nonumber \\ && \left. \frac{5}{9}\mu_d  \right) \sin^2\theta
- \frac{8}{9}  \left(\mu_u -\mu_d \right) \sin\theta \cos\theta\,,  \eea \bea
\mu^S_{{\rm sea}}(S_{11}^+(1535)) = \frac{a}{9}\left[ \left(5 + 2
\alpha^2 + \frac{2 \beta^2}{3} + \frac{4 \zeta^2}{3} \right) \mu_u+
\right. \nonumber \\  \left. \left( 4 + \alpha^2 + \frac{\beta^2}{3} + \frac{2 \zeta^2}{3}
\right)\mu_d + 3\alpha^2 \mu_s \right]  \cos^2\theta-   \nonumber \\
 \frac{5 a}{9} \bigg[ \left(5 + 2 \alpha^2 + \frac{2 \beta^2}{3} +
\frac{4 \zeta^2}{3} \right) \mu_u
+ \nonumber \\   \left( 4 + \alpha^2 +\frac{\beta^2}{3} + \frac{2 \zeta^2}{3}
\right)\mu_d +3 \alpha^2 \mu_s \bigg] \sin^2 \theta  \nonumber \\
+ \frac{8a}{9}  \left(1 + \alpha^2 + \frac{\beta^2}{3}
 + \frac{2 \zeta^2}{3}\right)(\mu_u -\mu_d) \sin\theta \cos\theta  \,, \eea  \bea
\mu^S_{{\rm orbit}}(S_{11}^+(1535)) &=& -\left(\frac{2}{9}\mu
(u_\uparrow  \rightarrow)
+\frac{1}{9}\mu (d_\uparrow \rightarrow) \right) \cos^2 \theta + \nonumber \\ &&
\left(\frac{10}{9}\mu (u_\uparrow  \rightarrow) +\frac{5}{9}\mu
(d_\uparrow  \rightarrow) \right)\sin^2\theta- \nonumber \\&&
 \frac{8}{9} \left( \mu (u_\uparrow  \rightarrow)
-\mu (d_\uparrow \rightarrow) \right) \sin\theta \cos\theta\,. \eea

The orbital angular momentum contribution $\mu^L$ to the magnetic
moment of a given baryon receives contributions from
the valence and sea quarks as \be \mu^L_{B} = \mu^L_{{\rm val}} +
\mu^L_{{\rm sea}} \,,\label{totmag2} \ee where  $\mu^L_{{\rm val}}$
and $\mu^L_{{\rm sea}}$ represent the contributions of the valence and
sea quarks to the magnetic moments due to the orbital angular momentum
polarizations. The details
of the valence quark calculations have already been presented in the
previous section,
whereas the sea quark spin polarizations corresponding to each baryon can be
calculated by substituting for each valence quark  with the third component of the
orbital angular momentum $\pm1$  \be
q^{(\pm1)}
\rightarrow  -T_{[q, ~GB]} q^{(\pm1)}+ |\psi(q^{(\pm1)})|^2\,, \label{qp2} \ee
 where $T_{[q, ~GB]}$ is the probability of emission of GBs
from a quark $q^{(\pm1)}$ and
$|\psi(q^{(\pm1)})|^2$ is the probability
of transforming a $q^{(\pm1)}$ quark, given as
\bea |\psi(u^{(\pm1)})|^2 &=& a d^{(\pm1)} + a \alpha^2 s^{(\pm1)} \,,
\nonumber \\|\psi(d^{(\pm1)})|^2 &=& a u^{(\pm1)} + a \alpha^2
s^{(\pm1)}\,, \nonumber \\ |\psi(s^{(\pm1)})|^2 &=& a \alpha^2
u^{(\pm1)} + a \alpha^2 d^{(\pm1)} \,. \eea
One  can calculate the valence and sea contributions of the orbital
angular momentum to the magnetic moment of the
baryons. As an example, the $\mu^L$ contribution to the magnetic
moment of $S_{11}^+(1535)$ is given as  \bea
\mu^L_{{\rm val}}(S_{11}^+(1535)) &=& \left(\frac{4}{9}\mu_u
+\frac{2}{9}\mu_d \right)
\cos^2\theta - \nonumber \\ && \left(\frac{1}{9}\mu_u + \frac{2}{9}\mu_d \right)
\sin^2\theta \,, \nonumber \\ \eea \bea
\mu^L_{{\rm sea}}(S_{11}^+(1535)) &=&  \frac{2a}{9}\left [-\left(1+2
\alpha^2 \right) \mu_u + \left(1- \alpha^2 \right)  \mu_d + \right. \nonumber \\ &&  \left. 3 \alpha^2
\mu_s \right] \cos^2 \theta +
\frac{a}{9}\left [-\left(1- \alpha^2 \right) \mu_u + \right. \nonumber \\ &&  \left.  \left(1+ 2
\alpha^2 \right)  \mu_d - 3\alpha^2 \mu_s \right] \sin^2 \theta\,.
\eea
The magnetic moments of the other $N^*$ resonances $S_{11}^0(1535)$,
$S_{11}^+(1650)$, and $S_{11}^0(1650)$ can also be calculated in
$\chi$CQM using similar methodology. The calculations have also been extended for the
magnetic moments of the first excited states (with $J^P
=\frac{1}{2}^-$) of the other baryons  of the ${\frac{1}{2}}^+$ octet.
All these results are discussed in detail in the next
section.

\section{Results and Discussion}
\label{result}
In this section, we first discuss the various input parameters needed for
the numeric calculation of the magnetic moment for the low-lying ${\frac{1}{2}}^-$ baryons. The calculation of the magnetic moments in the $\chi$CQM with {\it SU}(3)
broken symmetry
requires the symmetry breaking parameters $a$, $a\alpha^2$, $a
\beta^2$, and $a\zeta^2$, representing, respectively, the probabilities
of fluctuations of a constituent quark into pions, kaons, $\eta$, and
$\eta^{'}$. The best fit
to the set of parameters obtained by carrying out a fine grained
analysis of the spin and flavor distribution functions of the proton
\cite{chengmag1,chengmag2,chengmag3,hd1,hd2,hd3,hd4,hd5,nscharm} gives $$ a = 0.12 \,,~~ \alpha =
\beta = 0.45 \,, ~~\zeta = -0.15 \,.$$ In addition to the parameters of the $\chi$CQM, the contributions to the orbital angular momentum due to the quark sea are
characterized by the quark and GB masses. For the
constituent quark masses $u$, $d$, $s$, we have used their widely
accepted values in the hadron spectroscopy $M_u = M_d = 0.33$ GeV, and
$M_s = 0.51$ GeV.

Using the present formalism, we have calculated the magnetic moments of the negative parity $N^*$
resonances $S_{11}^+(1535)$, $S_{11}^0(1535)$,
$S_{11}^+(1650)$, and $S_{11}^0(1650)$ states and the results are presented in Table \ref{mag1}. For the sake of comparison, we
have also presented the results of $\chi$CQM of Ref. ~\cite{jliu}.

\begin{table*}
\caption{Magnetic moments of the low-lying $N^*$ resonances with $J={\frac{1}{2}}$ (in units of nuclear magneton)}\label{mag1}
\centering
\begin{tabular*}{\textwidth}{@{\extracolsep{\fill}}|ccr|rrrr|rrr|r|r|@{}} \hline
Baryons & Mass&NCQM & \multicolumn{8}{c|}{$\chi$CQM} & $\chi$CQM \\\cline{4-11}
& (MeV) & & \multicolumn{4}{c|}{$\mu^S$}&\multicolumn{3}{c|}{$\mu^L$}&$\mu_B (=\mu^S_B+\mu^L_B)$ & \cite{jliu}\\\cline{4-10}
 & & &$\mu^S_{{\rm val}}$ & $\mu^S_{{\rm sea}}$ & $\mu^S_{{\rm orbit}}$ & $\mu^S_B$ & $\mu^L_{{\rm val}}$ & $\mu^L_{{\rm sea}}$ & $\mu^L_B$ & & \\ \hline
$S_{11}^+$ & 1535&1.894& 1.411 & $-0.249$ & 0.530 & 1.692 &0.483& $-0.090$ & 0.393 & 2.085& 1.4\\
$S_{11}^0$ &1535&$-1.284$& $-1.192$ &0.167& $-0.519$ &$-$1.545 &$-0.092$ &0.067&$-$0.025& $-$1.570&$-$0.9\\
$S_{11}^+$  &1650&0.106 &$-$0.078& $-$0.108&$-$0.229 &$-$0.414 &0.184&$-$0.057&0.128 &$-$0.286& 0.1\\
$S_{11}^0$ &1650&0.951&1.192&$-$0.312&0.283&1.165 &$-$0.242&0.060 &$-$ 0.181 &0.984& 0.6 \\ \hline
\end{tabular*}
\end{table*}
In Table \ref{mag2},
we give the magnetic moments found in the present work for the low-lying  ${\frac{1}{2}}^-$ octet baryon resonances.
The explicit results for the
valence, sea, and orbital contributions of the spin part and
the valence and sea contributions of the orbital angular momentum
part of the magnetic moments are shown in these tables. For the sake of completeness,
we have also given the values found in the NCQM, obtained by summing the
contributions from the spin and orbital angular momentum of the valence quarks.

A cursory look at Table \ref{mag1} reveals that the magnitude of  the magnetic
moment is higher in the $\chi$CQM when compared to the NCQM predictions.
In general, the valence structure of the quarks dominates. However,
the presence of the sea quarks augments the absolute value of the magnetic moment,
especially in case of  $S^+_{11}(1650)$.
% as compared to the valence quarks contribution.

\begin{table*}
\caption{Magnetic moments of the first excited states (with  $J^P={\frac{1}{2}}^-$)
of the octet baryons (in units of nuclear magneton) }\label{mag2}
\begin{tabular*}{\textwidth}{@{\extracolsep{\fill}}|ccr|rrrr|rrr|r|@{}} \hline
Baryons & Mass &NCQM & \multicolumn{8}{c}{$\chi$CQM}  \\\cline{4-11}
& (MeV) & & \multicolumn{4}{c|}{$\mu^S(B)$}&\multicolumn{3}{c|}{$\mu^L(B)$}&$\mu_B(=\mu^S_B+\mu^L_B)$  \\\cline{4-10}
& && $\mu^S_{{\rm val}}$ & $\mu^S_{{\rm sea}}$ & $\mu^S_{{\rm orbit}}$ &$\mu^S_B$& $\mu^L_{{\rm val}}$ & $\mu^L_{{\rm sea}}$  &$\mu^L_B$&  \\ \hline
$p^*$ &1535 &1.894 &1.411 &$-0.249$& 0.530 &1.692 & 0.483&$-0.090$&0.393 & 2.085 \\
$n^*$ &1535& $-1.284$ & $-1.192$ &0.167& $-0.519$&$-$1.544 &$-0.092$ &0.067&$-$0.025 & $-$1.569  \\
$\Sigma^{*+}$& 1620& 1.814 & 1.297 & $-0.242$&0.349 &1.404& 0.518&$-0.122$&0.396& 1.800 \\
$\Sigma^{*-}$&1620&$-$0.689& $-0.333$&0.018&$-0.453$ &$-$0.768 &$-0.355$&0.117&$-$0.239& $-$1.007\\
$\Sigma^{*0}$&1620& 0.820&0.739 &$-$0.112&0.086 &0.713&0.081&$-$0.003&0.078& 0.791 \\
$\Xi^{*-}$&1690& $-$0.315& $-0.027$& $0.017$&$0.078$ &0.068 &$-0.288$& 0.055& $-$0.233 &$-$0.165\\
$\Xi^{*0}$& 1690&$-$0.990& $-$1.00& $-$0.209& $-$0.217 & $-$1.426 &0.011&$-$0.027& $-$0.016&$-$1.442 \\\hline
\end{tabular*}
\end{table*}

%\begin{table*}
%\centering
%\caption{Magnetic moments of the first excited states (with  $J^P={\frac{1}{2}}^-$) of the octet baryons (in units of nuclear magneton) }\label{mag2}
%\begin{tabular*}{\textwidth}{@{\extracolsep{\fill}}|ccr|rrrr|rrr|r|r|@{}} \hline
%Baryons & Mass &NCQM & \multicolumn{8}{c|}{$\chi$CQM} & Lattice \\\cline{4-11}
%& (MeV) & & \multicolumn{4}{c|}{$\mu^S$}&\multicolumn{3}{c|}{$\mu^L$}&$\mu_B(=\mu^S_B+\mu^L_B)$  & QCD \\\cline{4-10}
%& && $\mu^S_{{\rm val}}$ & $\mu^S_{{\rm sea}}$ & $\mu^S_{{\rm orbit}}$ &$\mu^S_B$& $\mu^L_{{\rm val}}$ & $\mu^L_{{\rm sea}}$  &$\mu^L_B$& &  \cite{lattice}\\ \hline
%$p^*$ &1535 &1.894 &1.411 &$-0.249$& 0.530 &1.692 & 0.483&$-0.090$&0.393 & 2.085 & $-$1.8\\
%$n^*$ &1535& $-1.284$ & $-1.192$ &0.167& $-0.519$&$-$1.544 &$-0.092$ &0.067&$-$0.025 & $-$1.569& $-$1.0 \\
%$\Sigma^{*+}$& 1620& 1.814 & 1.297 & $-0.242$&0.349 &1.404& 0.518&$-0.122$&0.396& 1.800&$-$0.6 \\
%$\Sigma^{*-}$&1620&$-$0.689& $-0.333$&0.018&$-0.453$ &$-$0.768 &$-0.355$&0.117&$-$0.239& $-$1.007& 1.0\\
%$\Sigma^{*0}$&1620& 0.820&0.739 &$-$0.112&0.086 &0.713&0.081&$-$0.003&0.078& 0.791& 0.1\\
%$\Xi^{*-}$&1690& $-$0.315& $-0.027$& $0.017$&$0.078$ &0.068 &$-0.288$& 0.055& $-$0.233 &$-$0.165&  0.8\\
%$\Xi^{*0}$& 1690&$-$0.990& $-$1.00& $-$0.209& $-$0.217 & $-$1.426 &0.011&$-$0.027& $-$0.016&$-$1.442&$-$0.5 \\ \hline
%\end{tabular*}
%\end{table*}

The magnetic
moments of $\Sigma^{*}$(1620) and $\Xi^{*}$(1690) are shown in
Table~\ref{mag2}, where the numbers for $N^*$(1535) are also displayed with the idea of showing the magnetic moments
of all the  ${\frac{1}{2}}^-$ octet  resonances together.
Table~\ref{mag2} shows that, just like in case of the $N^*$'s, the valence structure of the quarks dominates but the absolute value of the magnetic moment in $\chi$CQM  increases for the ${\frac{1}{2}}^-$ octet  resonances except for $\Sigma^{*+}$, $\Sigma^{*0}$, and $\Xi^{*-}$, where it decreases.

For the low-lying spin $\frac{1}{2}^{-}$ baryons,  so far there is no experimental information available
for the magnetic moments but we can compare our findings with those obtained within other models. Let us discuss the $N^*$ resonances first, out of which, for the $S_{11}$ (1535) an experiment is already being planned to measure its magnetic moment.  Several theoretical calculations of the magnetic moment of  $N^*(1535)$ have also been done earlier:  within CQM \cite{chiang},  $\chi$CQM ~\cite{jliu} and in the models based on hadronic degrees of freedom \cite{hyodo}, in which it has been found to arise due to pseudoscalar-baryon coupled channel dynamics.  The difference between the present work and that of Ref.~\cite{jliu} is the inclusion of the contribution to the magnetic moment arising from the  orbital
angular momentum of the sea quarks gained by the GB emission. This contribution is calculated through the generalized Cheng-Li
mechanism.  Our CQM results for the $N^*$'s coincide with those obtained in  Ref.~\cite{chiang}, but our $\chi$CQM results  are different with respect to Ref.~\cite{jliu} due to the absence of $\mu_{{\rm orbit}}^S$ contribution in the work of Ref.~\cite{jliu}. It can be seen from Table~\ref{mag1} that $\mu_{{\rm orbit}}^S$ gives a significant contribution to the total magnetic moment of baryons.

We now compare our results for $N^*$(1535) with those obtained in the study based on  chiral effective field theory \cite{hyodo}. In the present model the resonances have a three (valence) quark structure with one of the quarks excited to the P-wave and the consideration of the chiral symmetry takes into account the probability of the production of  Goldstone bosons. Although the hadronic models are also based on similar grounds, i.e., the chiral symmetry  and its spontaneous breaking, the structure of the baryon resonances in the two cases is different. In  Ref.~\cite{hyodo}, the magnetic moments of the $n^* (1535)$ and $p^*(1535)$ were found to be $-0.25 \mu_N$ and $+1.1 \mu_N$, respectively.  As can be seen, these results, although qualitatively similar, are quantitatively  different from those  listed in Table~\ref{mag1} of the present work.  Although, it should be mentioned that the simple one channel meson-baryon molecular picture of Ref.~\cite{chiang} yields $-0.56 \mu
 _N$ and $+1.86 \mu_N$ for the magnetic moments of $n^* (1535)$ and $p^*(1535)$, respectively, out of which at least $\mu_{p^*}$ is in better agreement with the quark model calculations.
Ideally, the wave function of the baryon resonances should be treated as an admixture of different configurations, but it is possible that one of those configuration dominates. If any experimental information is made available on the magnetic moments of baryon resonances, then a comparison of  the results found in quark models and hadronic models can shed light on the structure of the low-lying resonances. The experimental investigations on the magnetic moments of the $S_{11}(1535)$ resonance are, thus, most
welcome.

Going over to the discussions of the other octet resonances,
$\Sigma^*$(1620) is a controversial state, the results for this resonance obtained from the  partial wave analysis and the production experiments have been kept separately in Ref.~\cite{pdg}, mentioning that it is not clear if more than one resonance contribute to its signal \cite{pdg}. Models based on  effective field theories find  two different  $\Sigma^*$ (1620) with different spin parities. In Ref.~\cite{eo}, a resonance is found around 1620 MeV with spin-parity ${\frac{1}{2}}^-$ in the $K \Xi$ and coupled channel system.  On the other hand, a ${\frac{1}{2}}^+$ spin-parity  $\Sigma$ resonance with mass 1620 MeV is found to get dynamically generated in the $\pi \pi \Sigma$ system when the $\pi \Sigma$ subsystem forms $\Lambda$(1405) \cite{ninepr}.  Once again, measurement of the magnetic moment of this resonance can be useful in understanding it's nature. The known width of this resonance is $\sim$ 100 MeV, which is similar to the cases of $\Delta$(1232) and $N^*$(1535).  In case of $\Sigma$
 (1620), however, kaon beams would be required. Such a facility is available in JPARC, SPring-8, etc..

In case of ${\frac{1}{2}}^-$ $\Xi^*$ resonance, not much information is available. However, the CLAS Collaboration at the JLAB has recently initiated a cascade-physics program \cite{jlab2}, and  strangeness -2, -1  studies are  getting a special attention at KEK \cite{kek}.  We hope that the magnetic moments of the low-lying  $\Xi^*$'s will also get measured and our work will be useful in future.

We have compared our results also  with those available from the Lattice QCD calculations of Ref.~\cite{lattice}, where magnetic moments of the baryon resonances have been obtained from the mass shifts. The results of  Ref.~\cite{lattice} are very different, even by sign in most cases. For example the magnetic moment for the $p^*$ is found to be $-1.8$ $\mu_{N}$ in contrast with +1.8 $\mu_{N}$ found in the NCQM. However, it has  been mentioned in Ref.~\cite{lattice} that the signal for ${\frac{1}{2}}^-$ excited baryons in their calculations needs to be improved, together with other corrections to be checked. Results of an improved calculations are most awaited and should help in understanding the properties of the baryon resonances in future.

\section{Summary and conclusions}
\label{summary}
To summarize, using the chiral
constituent quark model ($\chi$CQM), we have
carried out a detailed analysis  of the magnetic moments of the negative parity $N^*$
resonances $S_{11}^+(1535)$, $S_{11}^0(1535)$,
$S_{11}^+(1650)$, and $S_{11}^0(1650)$ and the other low-lying  baryon resonances with $J^P={\frac{1}{2}}^-$. Using the generally
accepted values of the quark masses and the parameters of the $\chi$CQM, the explicit
contributions coming from the spin and the orbital angular momentum have been calculated. We find that our results do not coincide well with those obtained in
hadronic models based on chiral effective field theories. This can be understood since, although the picture of the presence of the meson cloud in the two approaches is similar, the structure of the baryon resonances
are different. The preliminary results available on the magnetic moments from lattice QCD \cite{lattice} are also not compatible with our findings (neither with the results available from other models  \cite{chiang,hyodo,jliu}) but  these preliminary lattice results are based on relatively poor
statistics, as mentioned in Ref.~\cite{lattice}. Results from such studies with improved statistics will be very important to make more robust comparisons in future.

Any future experimental measurement would have important implications in understanding the complex structure of these resonances.

\section{Acknowledgements}
The authors would like to thank the discussions with Prof. Jun He. N.~S and H.~D would like to thank  DAE-BRNS (Ref No:
2010/37P/48/BRNS/1445) for the financial support. K.~P.~K and A.~M.~T acknowledge the financial support from the Brazilian funding agencies FAPESP and
CNPq.
%\end{acknowledgements}

\end{document}